\documentclass[a4paper,fleqn]{cas-dc}

\usepackage[numbers,sort&compress]{natbib} 
\usepackage{subcaption}
\usepackage{xcolor}
\usepackage{multirow}
\usepackage{mathtools}
\usepackage{tabularx}
\usepackage{adjustbox}
\usepackage{amsmath, amssymb}
\usepackage{array}
\usepackage{lastpage}
\usepackage{nccmath}
\usepackage{amsmath}
\usepackage{cleveref}
\usepackage{hyperref}

\newenvironment{notes}{}{}

\def\tsc#1{\csdef{#1}{\textsc{\lowercase{#1}}\xspace}}
\tsc{WGM}
\tsc{QE}
\tsc{EP}
\tsc{PMS}
\tsc{BEC}
\tsc{DE}
\begin{document}
\let\WriteBookmarks\relax
\def\floatpagepagefraction{1}
\def\textpagefraction{.001}

\title{Electronic properties governing the phase stability and elastic anisotropy of C14 and C15 Cr–Hf–Nb Laves phases}
\shorttitle{M. Díaz-Choque, C. Ferrazza-Schuch and L.T.F. Eleno}
\shortauthors{}

\title [mode = title]{}                      
   \affiliation[1]{organization={Computational Materials Science Group (ComputEEL/MatSci), Department of Materials Engineering, Lorena School of Engineering, University of São Paulo (EEL-DEMAR-USP)},
            addressline={Estrada Municipal Chiquito de Aquino 1000}, 
            city={Lorena},
            postcode={12602-810}, 
            state={S\~ao Paulo},
            country={Brazil}}
    \affiliation[2]{organization={Professional School of Mechanical and Electrical Engineering (EPIME), National Technological University of Lima South (UNTELS)},
            addressline={Pj Villa el Salvador Grau. 1 Sector 3 Sublt 3 Lote. Mz A Villa el Salvador}, 
            city={Lima},
            postcode={150142}, 
            state={Lima},
            country={Perú}}
\author[1,2]{Martin Díaz-Choque}
\author[1]{Cauã Ferrazza Schuch}
\author[1]{Luiz Tadeu Fernandes Eleno\corref{co}}
\ead{luizeleno@usp.br}  
\cortext[co]{Corresponding author}

\begin{abstract}
This study utilizes Density Functional Theory (DFT) to investigate the thermodynamic stability, elastic anisotropy, and electronic properties of C14 and C15 Laves phases within the Cr--Hf--Nb system. Both formation enthalpies and comprehensive elastic property analyses confirm the energetic and mechanical stability of the C14 (HfNb$_2$, HfCr$_2$, NbCr$_2$) and C15 (HfCr$_2$, NbCr$_2$) phases. Furthermore, the evaluation of elastic anisotropy reveals a descending order of HfCr$_2$ > NbCr$_2$ > HfNb$_2$ for the C14 phase, contrasting with NbCr$_2$ > HfNb$_2$ > HfCr$_2$ for the C15 phase. Finally, electronic structure and COHP analyses indicate that strong anti-bonding behavior near the Fermi level within the XM$_2$ M--M bonds acts as a primary destabilization mechanism for both of these Laves phases.
\end{abstract}

\maketitle

\section{Introduction}

\begin{figure*}
    \centering
    \begin{subfigure}[b]{0.48\textwidth}
    \centering
        \includegraphics[height=.25\textheight]{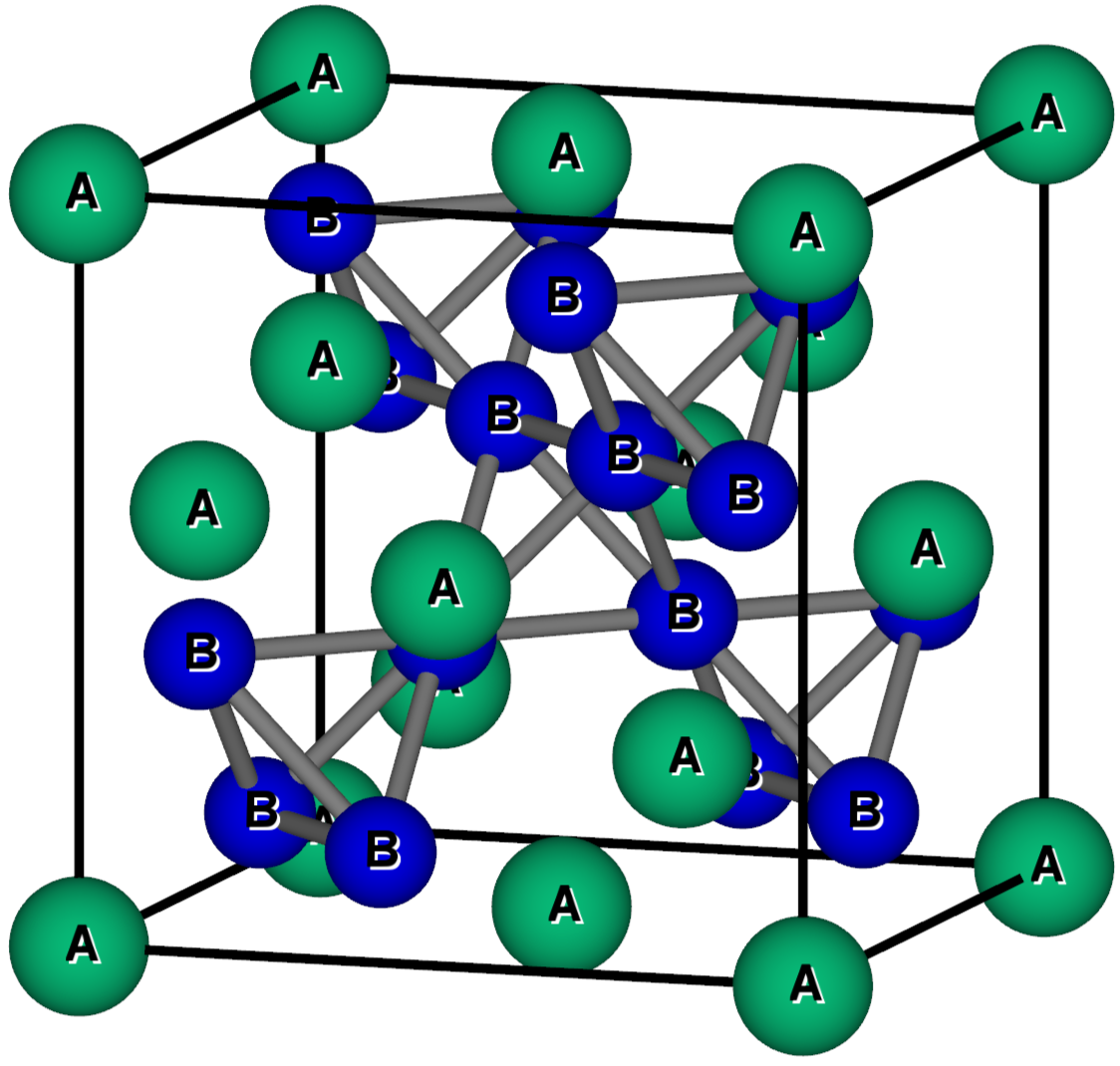}
        \caption{}
    \end{subfigure}
    \begin{subfigure}[b]{0.48\textwidth}
    \centering
        \includegraphics[height=.25\textheight]{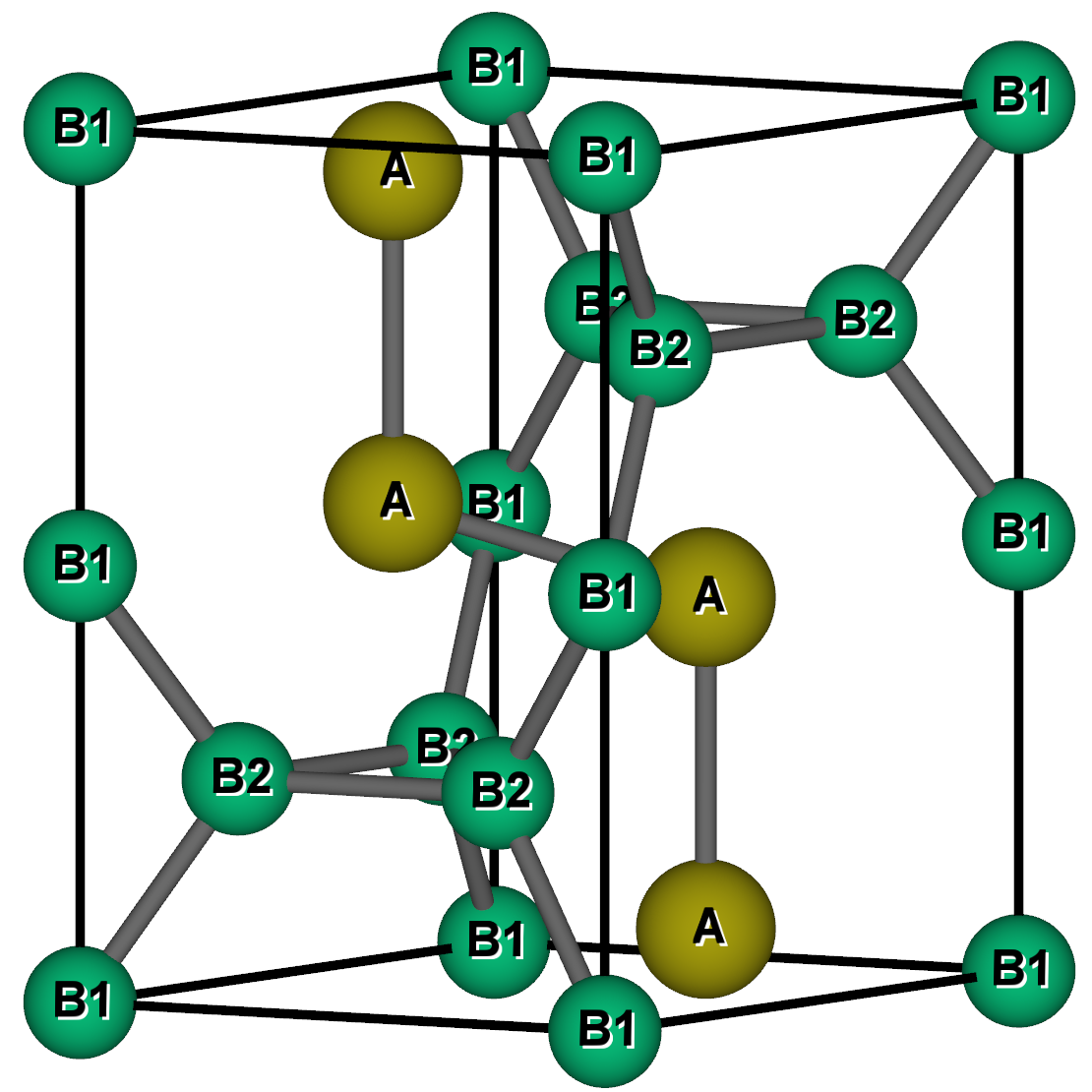}
        \caption{}
    \end{subfigure}
    \caption{Two Laves phase structures (AB$_2$): (a) C15  and (b) C14. Numbers indicate different Wyckoff sites.}
    \label{fig:c15-c14}
\end{figure*}

The Laves phase is a family of binary intermetallic compounds of composition $AB_2$ \cite{johnston1992struc}, crystallized in a face-centered cubic (FCC) lattice for C15 or in a hexagonal lattice (HCP) for C14 \cite{paul2004ab}. MgCu$_2$ (C15) crystallizes in the $Fd$$\bar{3}$$m$ ($227$ space group) where the Mg atoms form a cubic diamond net, while the Cu atoms form tetrahedral units (Fig. \hyperref[fig:c15-c14]{\ref{fig:c15-c14}(a)}), and MgZn$_2$ (C14) crystallizes in the hexagonal $P6_3/mmc$ ($194$ space group) where the Mg atoms form a hexagonal diamond net, and the Zn atoms are again described as forming tetrahedral units (Fig. \hyperref[fig:c15-c14]{\ref{fig:c15-c14}(b)}) \cite{roy2025structural}. They are potential high-temperature structural materials due to their high melting point, high strength and reasonably good oxidation resistance, and have long been considered for applications in the aerospace sector \cite{ma2014ab}. More recently, studies advancing quantum applications in C15 and C14, such as superconducting compounds \cite{ZHANG2024110448,ZHANG2025131122,Ma_2025,KOSHINUMA2022107643} and hydrogen storage \cite{YARTYS2022165219,PONSONI2022118317,doi:10.1021/acs.chemmater.5c01925} were also carried out. Various theoretical studies have been carried out for C15 and C14 XCr$_2$(X= Nb, Hf), including phase stability, elastic properties \cite{Chen2005ab, schmetterer2014new, YAO2007firt}. However, even the C15 Laves phases, are not yet widely exploited structural materials in industrial technology, due to the restricted mechanical properties, especially the brittle fracture at
low temperatures \cite{ma2014ab,YANG2012role}. Heaton and Samin \cite{heaton2024first} calculated the elastic constants and electronic structure of NbCr$_2$ C14 and C15 using molecular dynamics (MD). At the same time, they also estimated the Bulk modulus, Young’s modulus, and Poisson’s ratio. Hong and Fu \cite{hong1999phase} calculated the elastic constants and electronic structure of NbCr$_2$ using the full potential linearized augmented plane wave (FLAPW). For C15 NbCr$_2$, the elastic constants were higher than the experimental values. The calculated Young's moduli and Shear moduli were lower than the experimental values. On the other hand, Bai and Qin \cite{bai2023first} calculated the structural phase, elastic, and thermodynamic properties of HfCr$_2$ using VASP (PBE-GGA) in a pressure range of 0-200 GPa. From phonon spectra and elastic constants, HfCr$_2$ is mechanically stable in the C15 phase. Najrin et al. \cite{najrin2024comparative} calculated the elastic and anisotopic properties of the intermetallic compounds of the binary Laves phase C15 HfX$_2$ (X = Cr, Mo and W) using CASTEP (PBE-GGA). The composites showed excellent elastic stability, while HfMo$_2$ has shown higher ductility and machinability among all composites. On the other hand, the HfW$_2$ compound is relatively more anisotropic than HfCr$_2$ and HfMo$_2$. Finally, Liu et al. \cite{liu2017first} calculated the elastic property of NbCr$_2$  HfCr$_2$ using VASP with PW (PBE-GGA). NbCr$_2$ showed a higher bulk modulus and cauchy pressure, indicating a higher compressive strength and a more pronounced metallic character. In contrast, HfCr$_2$ showed a higher Shear modulus, Pugh ratio, and Young's modulus, but a lower Poisson's ratio.

In the present study, ab-intio calculations based on the density theory (DFT) of the C15 and C14 Lave phases in the binary systems Nb-Hf-Cr in the following compositions: HfNb$_2$, NbHf$_2$, HfCr$_2$, CrHf$_2$, NbCr$_2$, CrNb$_2$. Subsequently, The structural, elastic, anisotropic, electronic properties and bonding character of the stable compounds were analyzed.

\section{Methodology}

Calculations for the C14 and C15 structures were performed using the plane wave (PW) method \cite{hung2022quantum}, implemented in Quantum Espresso (QE) \cite{giannozzi2009quantum, giannozzi2017advanced}. This method, based on Density Functional Theory (DFT), addresses the many-body problem that involves interacting electrons and nuclei using the Kohn-Sham (KS) equations \cite{hohenberg1964inhomogeneous, kohn1965self}. Exchange and correlation interactions were described using the Generalized Gradient Approximation (GGA) with the Perdew-Burke-Ernzerhof (PBE) functional \cite{perdew1996generalized}, while electron nucleus interactions were used with ultra-soft GBRV pseudopotentials \cite{garrity2014pseudopotentials}. To ensure convergence in the calculations, a $50$ Ry plane wave energy cutoff, a $400$ Ry charge density cutoff and a $8\times8\times8$ k-point mesh with Marzari-Vanderbilt-DeVita-Payne cold smearing \cite{PhysRevLett.82.3296} of $0.01$ Ry were used for the Brillouin zone integrations using the Monkhorst-Pack scheme \cite{monkhorst1976special}. All numerical and structural parameters were optimized to guarantee a ground state convergence of $10^{-6}$ Ry in total energy and $10^{-4}$ Ry/a$_0$ (a$_0 \approx 0.529$ \AA) in total force acting on nuclei. We calculate the elastic properties using using Thermo\_PW  \cite{dalcorso_thermopw}, and for post-processing calculations we use VELAS \cite{Ran2023velas} for 2D and 3D graphical representations of anisotropy. 

For the electronic properties calculations, the electron-ion interaction was represented by the Blöchl’s projector-augmented wave (PAW) method \cite{PAW_BLOCHL}, with pseudopotentials from Pseudo Dojo library \cite{JOLLET20141246PAW}. A plane wave energy cutoff and the 120 Ry and 720 Ry charge density cutoff were used, respectively, with an increased k-point mesh of $16\times16\times16$ with tetrahedron method \cite{PhysRevB.49.16223} for DOS calculations. In order to analyze bond interaction, calculations of crystal orbital Hamilton populations (COHPs) and of the COHP integrated to the Fermi level (-iCOHP) were performed using the LOBSTER package \cite{Nelson2020LOBSTER,COHPdoi:10.1021/j100135a014}, and for post-processing calculations we use VESTA \cite{VESTA} for charge density in selected crystallographic planes.

\section{Results}

\subsection{Structural properties}

The Enthalpy of formation ($\Delta H$) is derived from the energy difference:
\begin{align}
\Delta H = \frac{E(F) - \left[N_x E(x^{E.F}) + N_y E(y^{E.F})\right]}{N_x + N_y}
\end{align}
where $E(F)$ is the total energy of the compound 'xy', $E(x^{E.F})$ and $E(y^{E.F})$ are the total energy of the atoms 'x' and 'y', respectively, $N_x$ and $N_y$ are the number of atoms 'x' and 'y', respectively. To derive physical parameters such as Bulk modulus, is fitted by the third-order Birch-Murnaghan equation of state is used \cite{birch1986equation}:
{\fontsize{9pt}{10pt}\selectfont
\begin{align*}
E_{\text{tot}}(a) = &E_0 + \frac{9 V_0 B_0}{16} \Bigg\{ 
\left[ \left( \frac{V_0}{V} \right)^2 - 1 \right]^3 B_0' \\
& + \left[ \left( \frac{V_0}{V} \right)^2 - 1 \right]^2 
\left[ 6 - 4 \left( \frac{V_0}{V} \right)^2 \right] 
\Bigg\}\tag{2}
\end{align*}
}
where $E_0$ is the energy of the ground state, $V_0$ is the altered volume, $B_0$ is the Bulk modulus, $B_0'$ is the pressure derivative of the bulk modulus and $V_0$ is the volume.

Table \ref{tab:C14-C15} shows our calculated results along with additional theoretical and experimental values. It can be seen that our calculation results agree well with the previous data. 

In the C$14$ Laves phase, the enthalpy of formation ($\Delta H$) of HfNb$_2$ has a value of -0.32 kJ/mol, while the compounds HfCr$_2$ and NbCr$_2$ in their NM (nomagnetic) and FM (ferromagnetic) states, present similar enthalpy values in each state, with values of -9.39 and -4.25 kJ/mol, respectively. On the other hand, the compounds NbHf$_2$, CrHf$_2$, and CrNb$_2$ show positive enthalpy, only, the enthalpy of CrHf$_2$ and CrNb$_2$ in their NM and FM states varies due to magnetism. The Bulk modulus (${B}$) of HfNb$_2$ is 147 GPa. The HfCr$_2$ and NbCr$_2$ in their NM and FM states show similar apparent modulus values of 189.82 and 227.93 GPa, respectively. The compounds CrHf$_2$ and CrNb$_2$ in their NM and FM states show a variation in the Bulk modulus due to magnetism.

In the C$15$ Laves phase, the $\Delta H$ of HfNb$_2$ has a value of 2.46 kJ/mol, while the compounds HfCr$_2$ in their NM and FM states shows a negative enthalpy, with minimal variation. In contrast, NbCr$_2$ in their NM and FM states shows a significant variation with values of -5.94 and -6.20 kJ/mol, respectively. On the other hand, compounds NbHf$_2$, CrHf$_2$, and CrNb$_2$ show positive enthalpy; only, the enthalpy of CrHf$_2$ and CrNb$_2$ in their NM and FM states varies due to magnetism. The ${B}$ of HfCr$_2$ in their FM and NM states maintains its value of $190.3$ GPa. The NbCr$_2$ in their NM and FM states shows minimal variation. Conversely, CrHf$_2$ and CrNb$_2$ in their NM and FM states show a variation in the Bulk modulus due to magnetism.

In the Laves phase C14 and C15, the magnetic moment ($\mu_{B}$) of the compounds HfNb$_2$, HfCr$_2$, and NbCr$_2$ in their NM and FM states show a value of zero, so they are NM compounds; unlike the compounds CrHf$_2$ and CrNb$_2$ that show magnetism in their FM state. Therefore, in the other sections we will show all the compounds in their NM state, in the supplementary section we will show all compounds in their FM state.
\begin{table*}[h]
\centering
\footnotesize   
\setlength{\tabcolsep}{4.5pt} 
\caption{Lattice parameters ($a$ and $c$) in \AA, enthalpy of formation ($\Delta H$) in kJ/mol, bulk modulus (${B}$) in GPa and magnetic moment ($\mu_{B}$) in Borh for the compounds HfNb$_2$, NbHf$_2$, HfCr$_2$, CrHf$_2$, NbCr$_2$, and CrNb$_2$ in C15 and C14 structures with NM and FM.}
\label{tab:C14-C15}
\begin{tabular}{lccccccccccc} 
\toprule
\multicolumn{12}{c}{\textit{C14 Structures}} \\
\midrule
\multicolumn{5}{c}{\textit{NM}} & \multicolumn{5}{c}{\textit{FM}} & \multicolumn{2}{c}{ \textit{Reference}} \\
Compound & $a$ (\AA) & $c$ (\AA) & $\Delta H$ (kJ/mol) & ${B}$ (GPa) & $a$ (\AA) & $c$ (\AA) & $\Delta H$ (kJ/mol)  & ${B}$ (GPa) & $\mu_{B}$ (Borh) \\
\midrule
HfNb$_2$ & 5.55 & 8.88  & -0.32 &  147.24& & & & && This work \\
NbHf$_2$ & 5.62 & 9.17  & 44.06 & 113.31& & & & &&  This work \\
HfCr$_2$ & 5.04 & 8.09  & -9.39 & 189.82& 5.04 & 8.09 &-9.39&189.82 & 0&  This work \\
              & 5.04\cite{Chen2005ab} & 8.09\cite{Chen2005ab}  & -9.86\cite{Chen2005ab} & 191\cite{Chen2005ab} & &  & & & &Other calcula. \\
              & 5.06\cite{sun2013first} & 8.14\cite{sun2013first}  & -10.74\cite{sun2013first} & &  & & & & &Other calcula.\\
              & 5.03\cite{pavluu2010thermodynamic} & 8.05\cite{pavluu2010thermodynamic}  & -8.70\cite{pavluu2010thermodynamic} & &  & & & & &Other calcula. \\
              & 5.06\cite{villars1991pearson} & 8.23\cite{villars1991pearson}  &       & &  & & & & &Exp.  \\
CrHf$_2$ & 5.55 & 8.78 & 99.18  & 105.05&5.62 & 9.46 & 79.62 &75.58 &13.79& This work \\
              & 5.52\cite{pavluu2010thermodynamic} & 8.65\cite{pavluu2010thermodynamic} & 96.63\cite{pavluu2010thermodynamic} & &  & & & & &Other calcula. \\
NbCr$_2$ & 4.89 & 8.07 & -4.25 &227.93 & 4.89 & 8.07 & -4.26 & 227.93 & 0& This work \\
              & 4.89\cite{lu2015thermodynamic} & 8.07\cite{lu2015thermodynamic}  & -2.58\cite{lu2015thermodynamic} & & & & & & &Other calcula. \\
              & 4.88\cite{hajra2023high} & 8.10\cite{hajra2023high}  &-2.60\cite{hajra2023high}& 219\cite{hajra2023high}  & & & & & &Other calcula. \\
              & 4.93\cite{villars1991pearson} & 8.12\cite{villars1991pearson}  &  & & & & & && Exp.\\
CrNb$_2$ & 5.36 & 8.32 & 70.51 &  156.64& 5.44 & 8.45 &61.71 &116.10& 11.61& This work \\
              & 5.36\cite{lu2015thermodynamic} & 8.32\cite{lu2015thermodynamic} & 7.23\cite{lu2015thermodynamic} & & & & & & &Other calcula.  \\
              & 5.38\cite{pavluu2010thermodynamic}  & 8.36\cite{pavluu2010thermodynamic}  & 76.52\cite{pavluu2010thermodynamic}  & & & & & & &Other calcula. \\
\midrule
\midrule
\multicolumn{12}{c}{\textit{C15 Structures}} \\
\midrule
\multicolumn{5}{c}{\textit{NM}} & \multicolumn{5}{c}{\textit{FM}} & \multicolumn{2}{c}{\textit{Reference}} \\
Compound & $a$ (\AA) & $c$ (\AA) & $\Delta H$ (kJ/mol) & ${B}$ (GPa) & $a$ (\AA) & $c$ (\AA) & $\Delta H$ (kJ/mol) & ${B}$ (GPa) & $\mu_{B}$ (Borh)  \\
\midrule
HfNb$_2$ & 7.99 & --     & 2.46  &  148.7& & & & &&  This work \\
NbHf$_2$ & 7.77 & --     & 50.56 & 101.3& & & & & & This work \\
HfCr$_2$ & 7.09 & --     & -11.10 &190.3 & 7.09 & -- &-11.06 & 190.3 &0&  This work \\
              & 7.08\cite{Chen2005ab} & --     & -11.63\cite{Chen2005ab} & & & && & &  Other calcula.  \\
              & 7.06\cite{pavluu2010thermodynamic} & --     & -10.38 \cite{pavluu2010thermodynamic} & & & & && &  Other calcula. \\
              & 7.14 \cite{villars1991pearson} & --     &      & & & & & &&  Exp.  \\
CrHf$_2$ & 7.82 & --     & 105.54 & 102.4 & 8.06 & -- & 80.43 & 74.2 &7.95&  This work \\
              & 7.77\cite{pavluu2010thermodynamic} & --    & 103.46\cite{pavluu2010thermodynamic} & & & & & & &  Other calcula.  \\
NbCr$_2$ & 6.94 & --     & -5.94 & 228.8 & 6.94 & -- & -6.20 & 228.0 &0&  This work \\
              & 6.94\cite{lu2015thermodynamic} & --     & -4.16\cite{lu2015thermodynamic} & & & & &- & & Other calcula.  \\
               & 6.82\cite{hong1999phase} & --     & -7.42\cite{hong1999phase} &252\cite{hong1999phase}& & & &- & & Other calcula.  \\
              & 6.93\cite{schmetterer2014new} & --     & -5.16\cite{schmetterer2014new} & -& 6.96\cite{schmetterer2014new}& -& -5.49\cite{schmetterer2014new}&- & & Other calcula.  \\
              & 6.99\cite{trojko1986structural}\cite{chu1994theoretical}& --     & -7.05\cite{martin1970thermodynamic} & 229 \cite{chu1994theoretical}  & & & & &  &Exp.\\
CrNb$_2$ & 7.46 & --    & 69.09  & 156.9 & 7.59 & --& 57.96 &120.0& 5.73 &  This work \\
& 7.50\cite{schmetterer2014new} & --     & 75.39 \cite{schmetterer2014new} &  & 7.50\cite{schmetterer2014new} & & 75.39\cite{schmetterer2014new}& &  &Other calcula.  \\
              & 7.45\cite{lu2015thermodynamic} & --     & 71.12 \cite{lu2015thermodynamic} &  & & & & &  &Other calcula.  \\
              & 7.50\cite{pavluu2010thermodynamic} & --     & 75.33\cite{pavluu2010thermodynamic} & & & & & &  &Other calcula.  \\
\bottomrule
\end{tabular}
\end{table*}

\subsection{Elastic properties}

\begin{table*}[H]
\centering
\footnotesize
\setlength{\tabcolsep}{4pt}
\caption{Elastic constants ($C_{11}$, $C_{12}$, $C_{13}$,$C_{33}$ $C_{44}$) in GPa, Young’s modulus ($E$) in GPa, bulk modulus ($B$) in GPa, shear modulus ($G$) in GPa, Pugh’s ratio ($B/G$), Poisson’s ratio ($\nu$), Debye temperature ($\theta_D$) in $K$, average sound velocity ($v_m$) in m/s, and hardness ($H_V$) in GPa for the compounds HfNb$_2$, NbHf$_2$, HfCr$_2$, CrHf$_2$, NbCr$_2$, and CrNb$_2$ in C14 structures with NM.}
\label{tab:elastic_c14}
\begin{tabular}{ccllllllllllllll}
\midrule
Compound & $C_{11}$ & $C_{12}$ & $C_{13}$ & $C_{33}$ & $C_{44}$ & $E$ & $B$ & $G$ & $B/G$ & $\nu$ & $\theta_D$ & $v_m$ & $H_V$ & Ref.\\
      NM   & (GPa) & (GPa) & (GPa) & (GPa) & (GPa) & (GPa) & (GPa) &(GPa) &&& ($K$) & (m/s) & (GPa) \\
\midrule
HfNb$_2$   &203.56 &124.02 &115.85 &210.83&33.88&106.54&146.63&38.63&3.80&0.38&241.81&2197.04&2.68 \\
NbHf$_2$&146.00&102.99&95.69&172.15&21.95&67.30&117.10&23.96&4.89&0.40&172.91&1600.632&1.44 \\
HfCr$_2$   &304.24&131.74&140.70&281.44&53.85&184.80&190.73&69.04&2.76&0.34&347.02&2866.96&5.81 \\
&302&139&148&275&51&161&194&59&3.29&0.36&324.8&2708.5&4.26&\cite{Chen2005ab}\\
CrHf$_2$   &134.05&95.96&85.22&146.89&29.00&68.85&105.32&24.75&4.26&0.39&182.02&1646.89&1.72 \\
NbCr$_2$   &336.29&176.11&173.15&341.47&59.12& 194.45& 228.89& 71.57&3.20&0.36&420.72&3403.12&5.05 \\
&304.15&189.93&164.84&336&73.11&-&220.38&68.17&3.23&-&-&-&4.81&\cite{sun2013structural}\\
CrNb$_2$&222.67&118.67&125.49&228.46&36.76&122.69&156.94&44.79&3.50&0.37&314.19&2726.79&3.26& \\
\midrule
\end{tabular}
\begin{notes}
\centering
\footnotesize
\text{\cite{Chen2005ab, sun2013structural} other calculations}
\end{notes}
\end{table*} 

\begin{table*}[H]
\centering
\footnotesize
\setlength{\tabcolsep}{6pt}
\caption{Elastic constants ($C_{11}$, $C_{12}$, $C_{44}$) in GPa, Young’s modulus ($E$) in GPa, bulk modulus ($B$) in GPa, shear modulus ($G$) in GPa, Pugh’s ratio ($B/G$), Poisson’s ratio ($\nu$), Debye temperature ($\theta_D$) in $K$, average sound velocity ($v_m$) in m/s, and hardness ($H_V$) in GPa for the compounds HfNb$_2$, NbHf$_2$, HfCr$_2$, CrHf$_2$, NbCr$_2$, and CrNb$_2$ in C15 structures with NM.}
\label{tab:C15-elastic}
\begin{tabular}{cclllllllllll}
\midrule
Compound & $C_{11}$ & $C_{12}$ & $C_{44}$ & $E$ & $B$ & $G$ & $B/G$ & $\nu$ & $\theta_D$ & $v_m$ & $H_V$ & Ref.\\
      NM   & (GPa) & (GPa) & (GPa) & (GPa) & (GPa) & (GPa) &  &  & ($K$) & (m/s) & (GPa) \\
\midrule
HfNb$_2$   & 139.87 & 100.27 & 22.73 & 60.69 & 113.47 & 21.51  & 5.27 & 0.41 & 183.48 & 1707.89 & 2.48 \\
NbHf$_2$   & 177.58 & 124.02 & 3.44 & 26.43 & 141.87  & 9.03  & 15.71 & 0.46 & 95.68 & 870.85 & 0.19 \\
HfCr$_2$   & 276.39 & 148.24 & 71.93  & 183.98 & 190.96 & 68.68 & 2.78 & 0.34 & 347.26 & 2867.30 & 5.75 \\
& 282 & 152 & 66 & 176 & 195& 65 & 3 & 0.35 & 315.4 & 2620.1 & 5.07& \cite{Chen2005ab} \\
& 292.6 & 151.62 & 79 & 200.97& 198.61& 75.48 & 2.63 & 0.33 & 364.40 & 2789.77& 6.48& \cite{najrin2024comparative} \\
& 252.17& 150.52 & 61.69 & 155.86& 184.40& 57.084 & 3.23 & 0.36 & 317 & 2600.29 & -& \cite{murad2024comprehensive} \\
CrHf$_2$   & 123.95 & 92.23 & 7.65  & 29.88 & 102.80 & 10.29  & 9.98 & 0.45 & 118.20 & 1076.50 & 0.35 \\
NbCr$_2$   & 310.66 & 188.46 & 76.34  & 190.17 & 229.20 & 69.83  & 3.28 & 0.36 & 416.04 & 3363.37 & 4.81 \\
 & 308 & 205 & 67  & - & 239 & 61  & 3.92 & 0.39 &- & -& 3.58 & \cite{li2009first} \\
 & 350.53 & 148.41 & 91.48  &  -& 215.78 & 95.31  & 2.26 & 0.31 &- & -& 9.15 & \cite{heaton2024first} \\
    & -& - & - & 214.8&192.74& 81.7&2.36& 0.32&453.9&3694.5& - & \cite{thoma1997elastic} \\
    & 303.7& 191.7 & 71.7&178.4& 229.3&65.1 &3.52&0.37&-& -&- &\cite{huang2021micromechanism}\\
        &335.59$ \pm $2.48& - &79.62$ \pm $0.02& 214.1& 229.4&79.6&-& 0.34&-&-&- & \cite{chu1995elastic} \\
CrNb$_2$   & 210.42 & 130.13 & 50.49  & 125.88 & 156.89 & 46.06 & 3.41 & 0.37 & 318.98 & 2772.03 & 3.44 \\
\midrule
\end{tabular}
\begin{notes}
\centering
\footnotesize
\text{\cite{Chen2005ab, najrin2024comparative, li2009first, huang2021micromechanism, murad2024comprehensive} other calculations, \cite{heaton2024first} molecular dynamics, \cite{thoma1997elastic, chu1995elastic} Exp.}
\end{notes}
\end{table*} 

These constants are obtained by calculations of the total energy under specific deformations that preserve the volume but modify the crystal symmetry~\cite{hichour2012theoretical}. To ensure mechanical stability in crystal lattices under atmospheric pressure ($P = 0$ GPa), the criteria proposed by Born~\cite{born1940stability} establish the following conditions:

for cubic
\begin{align*}
C_{11} - C_{12} > 0,\quad C_{11} + 2C_{12} > 0,\quad C_{44} >0\tag{3}
\end{align*}

for hexagonal
\begin{align*}
&C_{11} - C_{12} > 0,\quad 2C_{13}^2 < C_{33}(C_{11} + C_{12}), \\ &\quad C_{33}> 0,\quad C_{44} > 0,\quad C_{66} > 0\space \tag{4}
\end{align*}
Failure to meet any of these criteria leads to structural instability.

Based on these constants, mechanical properties such as the bulk modulus ($B_H$) and the shear modulus ($G_H$) can be calculated using the Voigt–Reuss–Hill approximations~\cite{reuss1929calculation, voigt1910lehrbuch, hill1952elastic}. 
\begin{align*}
B_H = \frac{B_{V} + B_{R}}{2},\ G_H = \frac{G_V + G_R}{2} \tag{5}
\end{align*}

From $B_H$ and $G_H$, the Young’s modulus ($E_H$) and Poisson’s ratio ($\nu$) can be derived using the equations:
\begin{align*}
E &= \frac{9B_HG_H}{3B_H + G_H},\ \nu=\frac{3B_H - 2G_H}{6B_H - 2G_H} \tag{6}
\end{align*}

The hardness ($H_V$) of a material provides insight into its elastic and plastic properties~\cite{qi2018electronic}. We can calculate it by means of the bulk modulus ($B$) and the shear modulus ($G$), expressed by the following equation~\cite{tian2012microscopic}:
\begin{align*}
H_V = 0.92 \left( \frac{G}{B} \right)^{1.137} G^{0.708} \tag{7}
\end{align*}

In addition, elastic constants and mechanical moduli make it possible to estimate acoustic properties such as the average sound velocity ($v_m$). These velocities are calculated using the following expressions~\cite{anderson1963simplified}:
\begin{align*}
v_m = \left[ \frac{1}{3} \left( \frac{2}{v_S^3} + \frac{1}{v_L^3} \right) \right]^{-1/3} \tag{8}
\end{align*}

where ($v_L$) is the longitudinal velocity and ($v_S$) is the shear velocity 
\begin{align*}
v_L = \sqrt{\frac{3B + 4G}{3\rho}}, \ v_S = \sqrt{\frac{G}{\rho}} \tag{9}
\end{align*}

The Debye temperature ($\theta_D$) is an important parameter of a solid. It is found in equations describing properties that arise from atomic vibrations and in theories involving phonons. we can calculate it from the average speed of sound ($v_m$)  by the following expression~\cite{anderson1963simplified}:
\begin{align*}
\theta_D = \frac{\hbar}{k_B} \left( \frac{6\pi^2 q}{V_0} \right)^{1/3} v_m \tag{10}
\end{align*}

Here, $q$ is the number of atoms in the unit cell, $V_0$ is the volume, and $\hbar$, $k_B$ are the reduced Planck constant and Boltzmann constant, respectively.

Table \ref{tab:elastic_c14} presents the calculated elastic properties for the C14 Laves phase in their NM state. The differences between the values obtained for the compounds HfCr$_2$ and NbCr$_2$ in their NM and FM (see Table 5 of the Supplementary Material) are small. Although experimental or theoretical data are not available for the elastic constants of HfNb$_2$, the calculated values for HfCr$_2$ and NbCr$_2$ agree well with theoretical data reported in the literature \cite{Chen2005ab, sun2013structural}.
The results indicate that the compounds HfNb$_2$, HfCr$_2$, and NbCr$_2$ exhibit good resistance to isotropic compression. Young's modulus ($E$) of NbCr$_2$ ($E = 194.45$ GPa) is the highest, followed by HfCr$_2$ and HfNb$_2$, suggesting that NbCr$_2$ is the stiffest. In relation to the Bulk modulus ($B$), NbCr$_2$ ($B = 228.89$ GPa) also exhibits the highest value, demonstrating greater resistance to volumetric deformation. The shear modulus ($G$) follows the same trend, with NbCr$_2$ ($G = 71.51$ GPa) outperforming the others, indicating stronger directional bonding. The order of the elastic moduli ($B$, $G$, and $E$) is as follows: NbCr$_2$ $>$ HfCr$_2$ $>$ HfNb$_2$.
According to Pugh’s criterion \cite{pugh1954magazine}, which uses the ratio of Bulk to Shear modulus ($B/G$) with a critical value of approximately 1.75 to distinguish between ductile and brittle behavior, all the compounds analyzed present values above this threshold, indicating ductile behavior. HfNb$_2$ ($B/G = 3.80$) shows the highest value, followed by NbCr$_2$ and HfCr$_2$, suggesting that HfNb$_2$ has the highest plasticity, while HfCr$_2$ is the most brittle. This trend is also supported by the Poisson’s ratio ($\nu$), where HfCr$_2$ ($\nu = 0.34$) shows the lowest value. The ductility order is as follows: HfNb$_2$ $>$ NbCr$_2$ $>$ HfCr$_2$.
Finally, the Debye temperature ($\theta_D$) and the average sound velocity ($v_m$) are highest for NbCr$_2$ ($\theta_D = 420.72$ K; $v_m = 3403.12$ m/s), followed by HfCr$_2$ and HfNb$_2$, indicating higher dynamic stiffness. However, Vickers hardness ($H_V$) shows a different behavior, with HfCr$_2$ ($H_V = 5.81$ GPa) showing the highest value, followed by NbCr$_2$ and HfNb$_2$. Until now, although there are no experimental or theoretical data to compare with these calculated results, the data would be useful for future research.
Table \ref{tab:C15-elastic} presents the calculated elastic properties for the C15 Laves phase in their NM state. The differences between the values obtained for the compounds HfCr$_2$ and NbCr$_2$ in their NM and FM (see Table 6 of the Supplementary Material) are small. The calculated results of HfCr$_2$ are in good agreement with other theoretical values \cite{Chen2005ab, najrin2024comparative, murad2024comprehensive}. The NbCr$_2$ NM is in good agreement with other theoretical \cite{heaton2024first,li2009first, huang2021micromechanism} and experimental \cite{thoma1997elastic, chu1995elastic} values. The order of the elastic moduli ($B$, $G$, and $E$) is as follows: NbCr$_2$ $>$ HfCr$_2$ $>$ HfNb$_2$. The order of ductility ($B/G$, $\nu$) is as follows: HfNb$_2$ $>$ NbCr$_2$ $>$ HfCr$_2$. Finally, $\theta_D$ and $v_m$ are highest for NbCr$_2$ ($\theta_D = 416.04$ K, $v_m = 3363.37$ m/s), followed by HfCr$_2$ and HfNb$_2$, indicating higher dynamic stiffness. However, $H_V$ shows a different behavior, with HfCr$_2$ ($H_V = 5.75$ GPa) showing the highest value, followed by NbCr$_2$ and HfNb$_2$.

\subsection{Anisotropic properties}
The evaluation of the elastic performance of materials involves the analysis of anisotropy, which describes the directional variation of their mechanical properties~\cite{tindibale2023elastic}. \\ Elastic anisotropy ($A$) influences mechanisms such as ani-
sotropic plastic deformation, microcrack propagation, and elastic instability~\cite{fang2019energy}. The shear anisotropy parameter can be calculated using the Zener anisotropy factor~\cite{zener1949elasticity}, defined as:
\begin{align*}
A = \frac{2C_{44}}{C_{11} + C_{12}} \tag{11}
\end{align*}

where $C_{11}$, $C_{12}$, and $C_{44}$ are the elastic constants of the crystal. For an isotropic crystal, $A = 1$; deviations from this value reflect the degree of anisotropy~\cite{tindibale2023elastic}. Materials with $A < 0$ violate Born’s criteria, classifying them as mechanically unstable~\cite{wu2019elastic}.

Likewise, the universal anisotropy index ($A^U$). A value of $A^U = 0$ indicates perfect isotropy, whereas deviations from zero reflect anisotropic behavior~\cite{Ranganathan2008unisersal}. This index is given by:
\begin{align*}
A^U = 5\frac{G_V}{G_R} - \frac{B_V}{B_R} - 6 \tag{12}
\end{align*}

To enable further verification, can be examined via the equivalent anisotropy index $A_{eq}$~\cite{Ranganathan2008unisersal}, defined as:
\begin{align*}
A_{\text{eq}} = \left(1 + \frac{5}{12}A^U\right) + \sqrt{\left(1 + \frac{5}{12}A^U\right)^2 - 1} \tag{13}
\end{align*}

Elastic anisotropy can also be evaluated through fractional indices associated with the bulk and shear moduli~\cite{Chung1967the}, defined by the following equations:
\begin{align*}
A^G = \frac{G_V - G_R}{G_V + G_R}, \ A^B = \frac{B_V - B_R}{B_V + B_R} \tag{14}
\end{align*}

These indices provide specific insights into distinct elastic parameters of the material. Values of $A^G = A^B = 0$ indicate perfect elastic isotropy, while $A^G = A^B = 1$ correspond to maximum elastic anisotropy~\cite{Ravindran1998density, Maxisch2006elastic}.

To characterize the anisotropy of each elastic modulus, a new anisotropy index \( A^X \) was proposed that was defined as follows\cite{ran2021phase}:
\begin{equation}
A^X =
\begin{cases}
\dfrac{X_{\text{std}}}{X_{\text{mean}}}, & \text{if } \text{sign}(X_{\text{max}}) = \text{sign}(X_{\text{min}}) \\
\infty, & \text{otherwise} \tag{15}
\end{cases}
\end{equation}

where \( X_{\text{std}} \) is the standard deviation of each elastic modulus \( X \), and \( X_{\text{mean}} \) is the average value of each elastic modulus \( X \).
\begin{table*}[H]
\footnotesize   
\setlength{\tabcolsep}{12pt}
\caption{Shear moduli of Voigt ($G_V$) and Reuss ($G_R$), Bulk modulus of Voigt ($B_V$) and Reuss ($B_R$), shear anisotropy index ($A^G$), bulk anisotropy index ($A^B$), universal anisotropy index ($A^U$), and equivalent anisotropy measure ($A_{\text{eq}}$) or the compounds HfNb$_2$, NbHf$_2$, HfCr$_2$, CrHf$_2$, NbCr$_2$, and CrNb$_2$ in C14 structures with NM.}
\label{tab:c14-anisotropy}
\begin{tabular}{cclllllllll}
\midrule
Compound & $G_V$ & $G_R$ &$B_V$&$B_R$& $A^G$ & $A^B$  & $A^U$ & $A_{\text{eq}}$ &Ref.\\
       NM     & (GPa) & (GPa) &     & &   &     &    &     &     \\
\midrule
HfNb$_2$ &38.99&38.28&147.71&145.71&0.0092&\ 0.0074&0.11&1.35& \\
NbHf$_2$&24.40&23.52&116.99&117.22&0.0183&-0.0010
&0.18&1.48& \\
HfCr$_2$   &70.58&67.50&190.69&190.77&0.0223&-0.0002&0.23&1.54& \\
CrHf$_2$   &-35.55&38.67&104.57&107.15&-23.80&-0.0012&-9.61&-0.17& \\
NbCr$_2$   &72.44&70.70&228.76&229.02&0.0122&-0.0001&0.12&1.37& \\
&68.98&67.37&220.39&220.37&-&-&-&-&\cite{sun2013structural} \\
CrNb$_2$   &45.38&44.19&157.01&156.87&0.0133&\ 0.0005&0.14&1.40& \\
\midrule
\end{tabular}
\begin{notes}
\centering
\footnotesize
\text{\cite{sun2013structural} other calculations}
\end{notes}
\end{table*}
\begin{table*}[H]
\setlength{\tabcolsep}{12pt}
\caption{Shear moduli of Voigt ($G_V$) and Reuss ($G_R$), Zener anisotropy factor ($A$), shear anisotropy index ($A^G$), universal anisotropy index ($A^U$), and equivalent anisotropy measure ($A_{\text{eq}}$) or the compounds HfNb$_2$, NbHf$_2$, HfCr$_2$, CrHf$_2$, NbCr$_2$, and CrNb$_2$ in C15 structures with NM.}
\label{tab:c15-anisotropy}
\begin{tabular}{ccllllllll}
\midrule
Compound & $G_V$ & $G_R$ & $A$ & $A^G$ & $A^U$ & $A_{\text{eq}}$ &Ref. \\
       NM    & (GPa) & (GPa) &   &   &    &     &   &   \\
\midrule
HfNb$_2$  & 21.56 & 21.46 & 1.15  & 0.002   & 0.02 & 1.15 &\\
NbHf$_2$   & 12.78 & 5.28 & 0.13 & 0.415   & 7.10 & 7.79& \\
HfCr$_2$  & 68.73 & 68.57  & 1.12  & 0.002  & 0.02  & 1.12 &\\
&-&-&1.007&-&-&-&\cite{Chen2005ab}\\
&75.59&75.36&1.12&0.002&0.02&1.15&\cite{najrin2024comparative} \\
&-&-&1.21&-&-&-&\cite{murad2024comprehensive} \\
CrHf$_2$  & 10.93 & 9.65  & 0.48   & 0.062  & 0.67   & 2.07 &\\
NbCr$_2$   & 70.25 & 69.41  & 1.25   & 0.006   & 0.06   & 1.25& \\
&-&-&1.26&-&-&-& \cite{huang2021micromechanism}\\
& 64.73 & 63.64  & 1.30 & 0.008& 0.08& 1.29& \cite{liu2023alloying}\\
CrNb$_2$   & 46.36 & 45.77  & 1.26   & 0.006  & 0.06   & 1.26 &\\
\midrule
\end{tabular}
\begin{notes}
\centering
\footnotesize
\text{\cite{Chen2005ab, najrin2024comparative, murad2024comprehensive, huang2021micromechanism, liu2023alloying} other calculations}
\end{notes}
\end{table*}

The calculated values of the anisotropy index $A$, $A^G$, $A^B$, $A^U$, $A_{\text{eq}}$ are the same for the FM (see Table 7, 8 of the Supplementary Material) and the NM part of all compounds. Table \ref{tab:c14-anisotropy} shows our calculated results for Laves C14 NM, it can be seen that the $A^B$ value of NbCr$_2$ is much lower than that of HfCr$_2$ and HfNb$_2$, indicating that the percentage of compression anisotropy of NbCr$_2$ is the lowest among the three compounds. However, the universal anisotropy index $A^U$, the equivalent anisotropy measure $A_{eq}$ and the shear anisotropy $A^G$ of HfCr$_2$ are higher than those of NbCr$_2$ and HfNb$_2$. It can be seen that HfCr$_2$ has the greatest anisotropy. The universal anisotropy index $A^U$ takes into account the contributions of $G$ and $B$. In general, the degree of anisotropy is compared on the basis of the value of $A^U$. Therefore, the order of the universal anisotropy index of C14-NM is: HfCr$_2$ $>$ NbCr$_2$ $>$ HfNb$_2$.
Table \ref{tab:c15-anisotropy} shows our calculated results for Laves C15 NM. The calculated results of HfCr$_2$ are in good agreement with other theoretical values \cite{Chen2005ab, najrin2024comparative, murad2024comprehensive}. The NbCr$_2$ NM is in good agreement with other theoretical \cite{liu2023alloying, huang2021micromechanism}. It can be seen that $A$, $A^G$, $A^U$ and $A_{eq}$ of NbCr$_2$ are higher than those of HfNb$_2$ and HfCr$_2$. It can be seen that NbCr$_2$ has the greatest anisotropy. The universal anisotropy index $A^U$ takes into account the contributions of $G$ and $B$, and that the values of $A$ equal $A_{eq}$. In general, the degree of anisotropy is compared on the basis of the value of $A^U$. Therefore, the order of the universal anisotropy index of C15-NM is: NbCr$_2$ $>$ HfNb$_2$ $>$ HfCr$_2$.

\begin{figure*}[!ht]
    \centering
        \includegraphics[width=\textwidth]{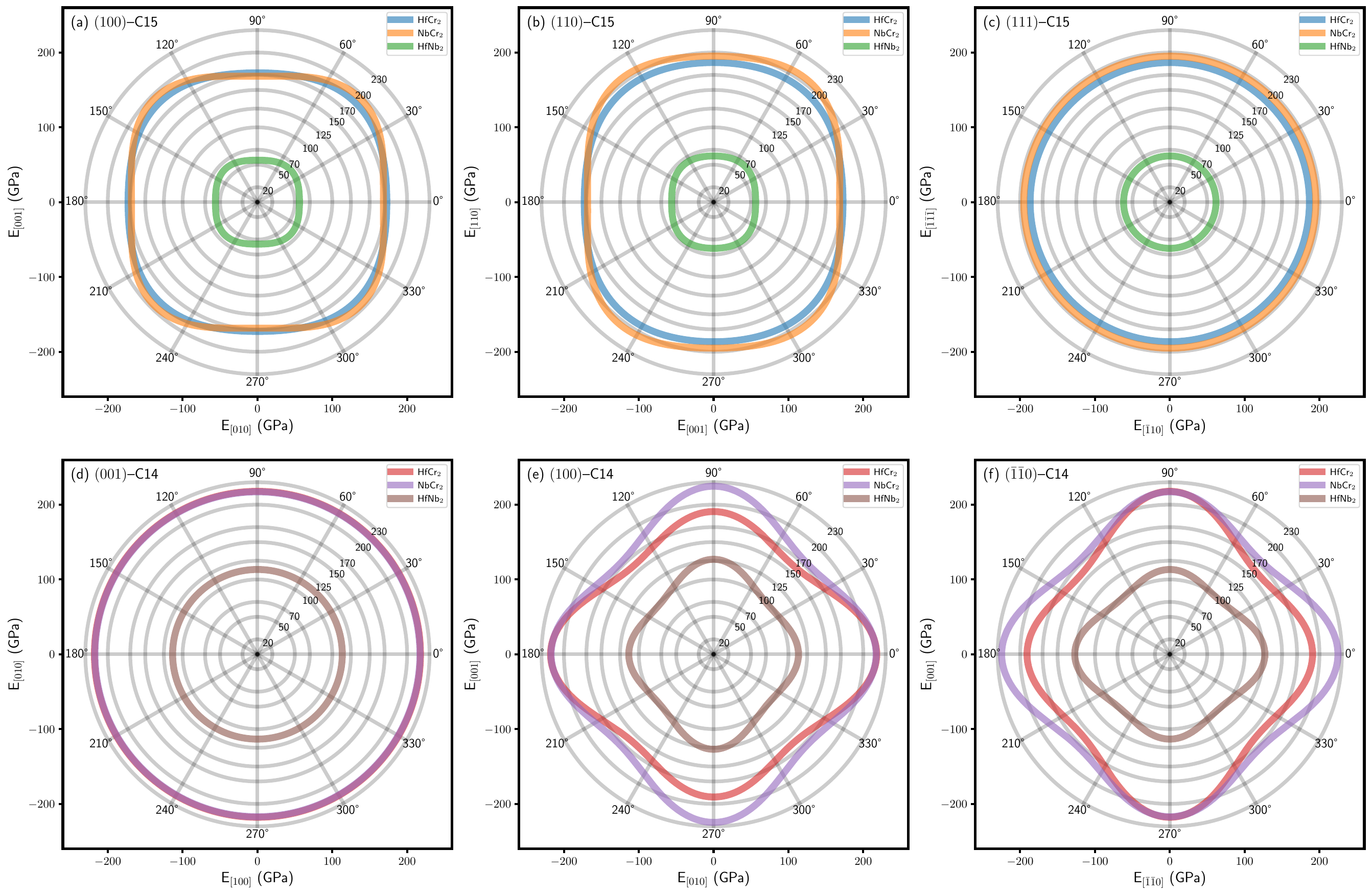}
        \caption{Young’s modulus $E$ (in GPa) for HfCr$_2$, NbCr$_2$, HfNb$_2$ in the C15 and C14 structures. Images  (a), (b), and (c) show the results in the (100), (110), and (111) planes, respectively, for the C15 structure; images (d), (e), and (f) show the results in the (001), (100), and ($\bar{1} \bar{1} 0$) planes, respectively, for the C14 structure.}
    \label{fig:C14-15}
\end{figure*}

The projection map of the surface of Young's modulus for the compounds HfCr$_2$, NbCr$_2$, and HfNb$_2$ with the structure of C14-NM is shown in Fig. \hyperref[fig:C14-15]{\ref{fig:C14-15}(d)-(f)}. It can be seen that the 3D surface structure of Young's modulus (see Fig. 4, 7, and 10 of the Supplementary Material) for these compounds deviates from a perfect sphere, indicating that Young's modulus is anisotropic. From a numerical perspective, the deformation resistance of NbCr$_2$ in the $(100)$ and ($\bar{1}$$\bar{1}$$0$) planes is higher than that of HfNb$_2$ in the same planes. It is also evident that, although HfNb$_2$ exhibits a good ability to resist deformation in the $(100)$ and ($\bar{1}$$\bar{1}$$0$) planes, this resistance is still lower than that of HfCr$_2$ in the same planes.
On the other hand, the anisotropy of HfCr$_2$ in the $(100)$ and ($\bar{1}$$\bar{1}$$0$) planes is more pronounced than that of HfNb$_2$. Although HfNb$_2$ also shows significant anisotropy in these planes, it remains lower than the anisotropy observed for NbCr$_2$.

The projection map of the surface of Young's modulus for the compounds HfCr$_2$, NbCr$_2$, and HfNb$_2$ with the structure of C15-NM is shown in Fig. \hyperref[fig:C14-15]{\ref{fig:C14-15}(a)-(c)}. It can be seen that the 3D surface structure of Young's modulus (see Fig. 13, 15, and 19 of the Supplementary Material) for these compounds deviates from a perfect sphere, indicating that Young's modulus is anisotropic. From a numerical perspective, the deformation resistance of NbCr$_2$ in the $(100)$ and ($\bar{1}\bar{1}0$) planes is higher than that of HfNb$_2$ in the same planes. It can also be noted that, although HfNb$_2$ shows good resistance to deformation in the $(100)$ and ($\bar{1}\bar{1}0$) planes, this resistance is still lower than that of HfCr$_2$ in the same directions.
On the other hand, the anisotropy of NbCr$_2$ in the $(100)$ and ($\bar{1}\bar{1}0$) planes is greater than that of HfCr$_2$. Although HfNb$_2$ also presents significant anisotropy in these planes, it remains lower than that observed in NbCr$_2$.
\subsection{Electronic structure and bonding analysis}

\begin{figure*}[!ht]
    \centering
        \includegraphics[width=\linewidth]{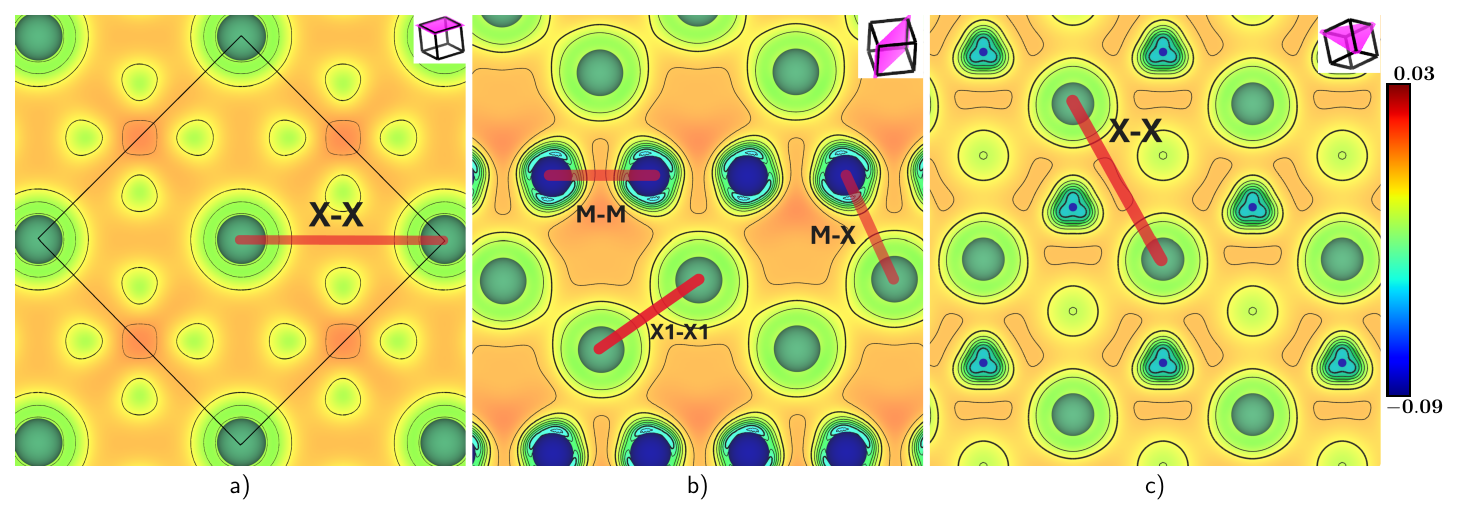}
        \includegraphics[width=\linewidth]{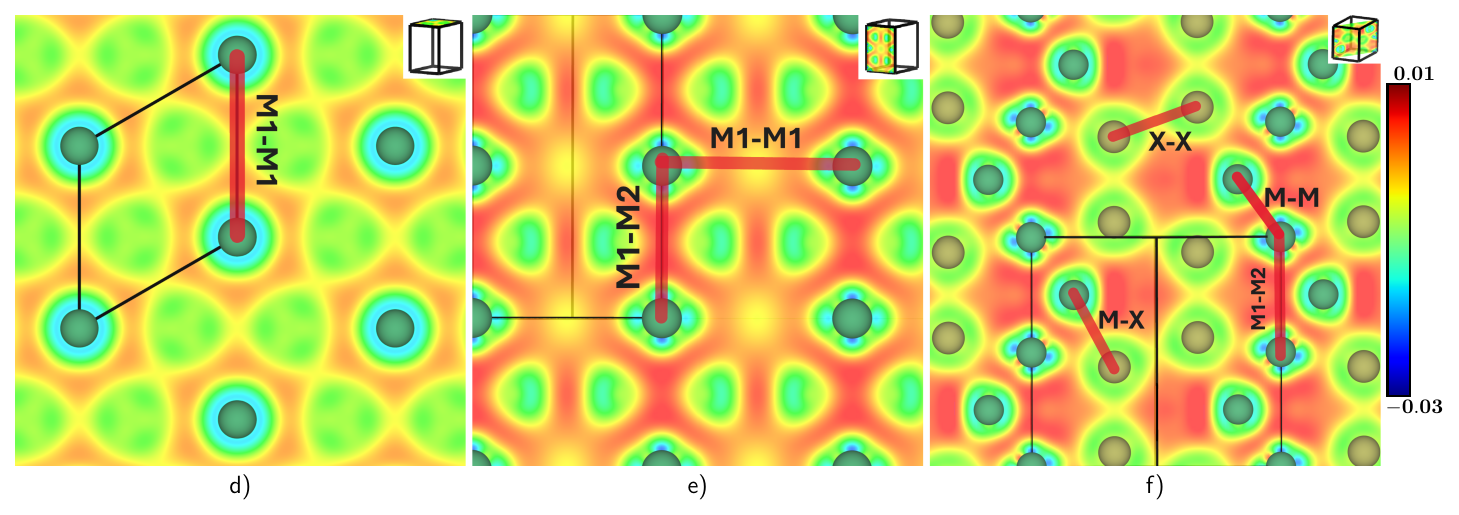}
        \caption{Charge density difference between the valence charge density of the crystal and the superposition of the valence charge density of the (a)--(c) NbCr$_2$--C15 and (d)--(f) HfNb$_2$--C14 constituent atoms. The crystallographic planes for C15 and C14 structures are, respectively: (a) $(100)$, (b) $(110)$ and (c) $(111)$; (d) $(001)$, (e) $(100)$ and (f) $(\bar{1}\bar{1}0)$. Cr, Nb and Hf atoms are represented as blue, green and yellow spheres, respectively. The color scales are in the units of e/Bohr$^3$.}
    \label{fig:CHG}
\end{figure*}
\begin{figure}[htb]
    \centering
    \includegraphics[width=0.98\columnwidth]{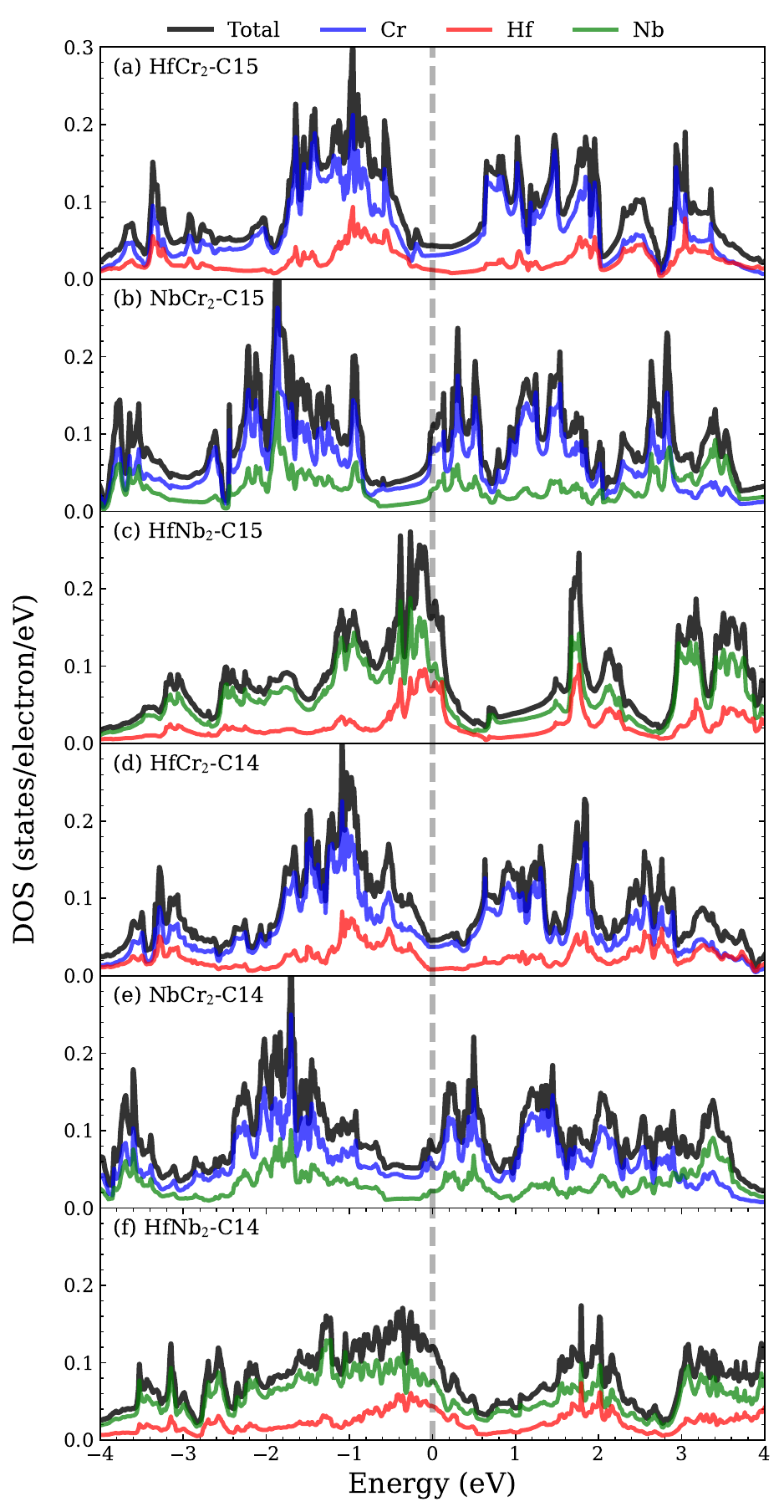}
    \caption{Total and partial density of states for (a)--(c) C15 and (d)--(f) C14 compounds at zero pressure. The Fermi level, set at 0, is indicated by the dashed vertical line.}
    \label{fig:DOS}
\end{figure}
\begin{figure}[htb]
    \centering
    \includegraphics[height=6.15cm, width=0.48\textwidth]{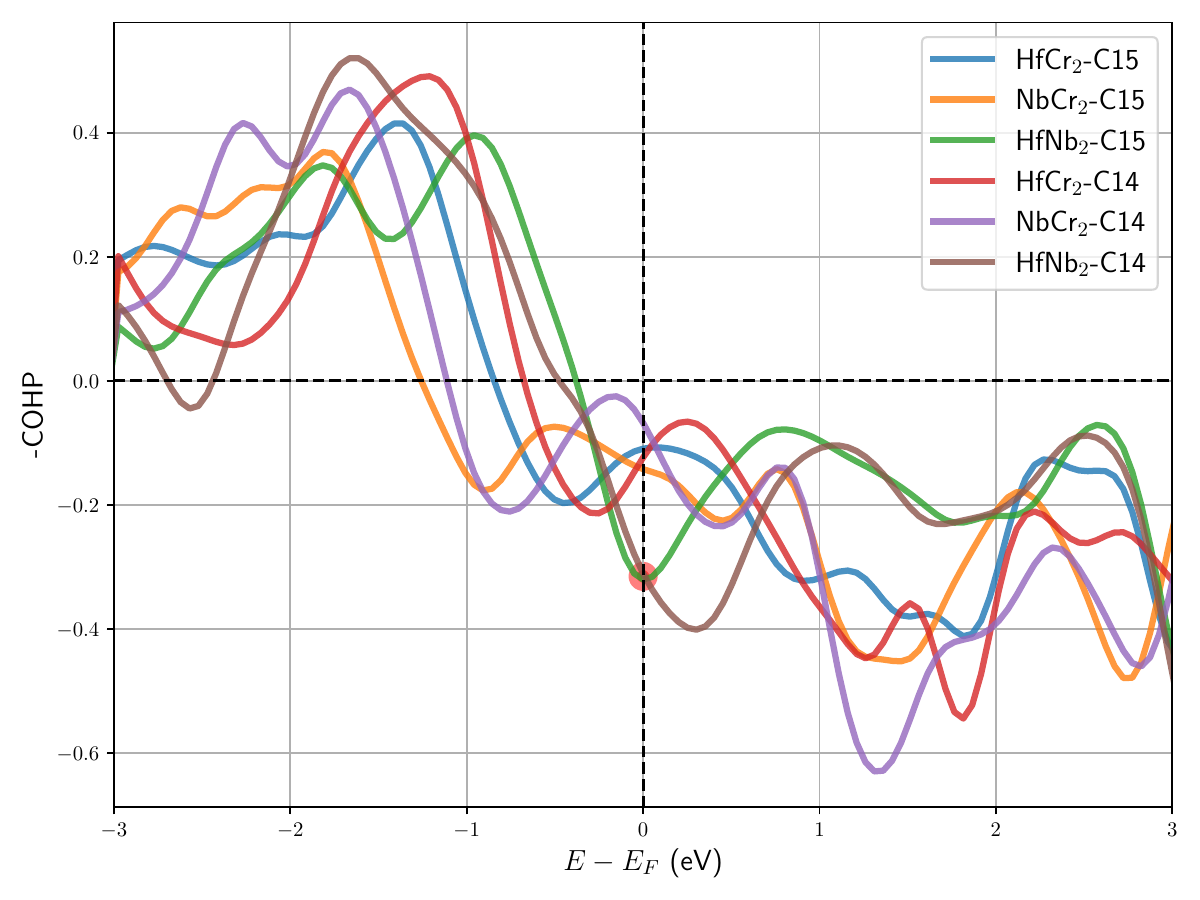}
    \caption{–COHPs plotted as a function of energy for the M--M bonds in C15 (Fig.\hyperref[fig:comparisons]{~\ref{fig:CHG}(b)}) and C14 (Fig.\hyperref[fig:comparisons]{~\ref{fig:CHG}(f)}) XM$_2$ compounds. The red point at $E=E_f$ highlights -COHP value for the two compounds with least stability.}
    \label{fig:COHP}
\end{figure}
\begin{table}[h!]
\centering
\caption{Bond lengths and ICOHP values (integral of the COHP up to the Fermi
energy, in eV) for C14 and C15 XM$_2$ compounds. An explicit depiction of the bond types is shown in Fig. \ref{fig:CHG}.}
\begin{tabular}{llcc}
\hline
Compound & Bond Type & Length (\AA) & ICOHP (eV) \\
\hline
\multirow{4}{*}{HfCr$_2$-C14} 
    & M--X       & 2.896 & $-1.140$ \\
    & M--M     & 2.492 & $-1.402$ \\
    & M1--M2     & 4.046 & $-0.093$ \\
    & M1--M1     & 5.048 & $-0.013$ \\
    & X--X       & 3.068 & $-1.395$ \\
\hline
\multirow{4}{*}{NbCr$_2$-C14} 
    & M--X       & 2.858 & $-1.105$ \\
    & M--M     & 2.404 & $-1.665$ \\
    & M1--M2     & 4.035 & $-0.104$ \\
    & M1--M1     & 4.897 & $-0.022$ \\
    & X--X       & 3.073 & $-1.197$ \\
\hline
\multirow{4}{*}{HfNb$_2$-C14} 
    & M--X       & 3.193 & $-1.183$ \\
    & M--M     & 2.693 & $-3.318$ \\
    & M1--M2     & 4.442 & $-0.231$ \\
    & M1--M1     & 5.556 & $-0.012$ \\
    & X--X       & 3.308 & $-1.144$ \\
\hline
\multirow{3}{*}{HfCr$_2$-C15} 
    & M--X       & 2.940 & $-1.068$ \\
    & M--M       & 2.507 & $-1.336$ \\
    & X1--X1     & 3.071 & $-1.356$ \\
    & X--X       & 5.014 & $-0.023$ \\
\hline
\multirow{3}{*}{NbCr$_2$-C15} 
    & M--X       & 2.878 & $-1.099$ \\
    & M--M       & 2.455 & $-1.409$ \\
    & X1--X1     & 3.006 & $-1.421$ \\
    & X--X       & 4.909 & $-0.034$ \\
\hline
\multirow{3}{*}{HfNb$_2$-C15} 
    & M--X       & 3.314 & $-0.992$ \\
    & M--M       & 2.826 & $-2.408$ \\
    & X1--X1     & 3.462 & $-0.909$ \\
    & X--X       & 5.653 & $-0.007$ \\
\hline
\label{ICOHP}
\end{tabular}
\end{table}
\begin{figure*}[!ht]
    \centering
        \includegraphics[width=\linewidth]{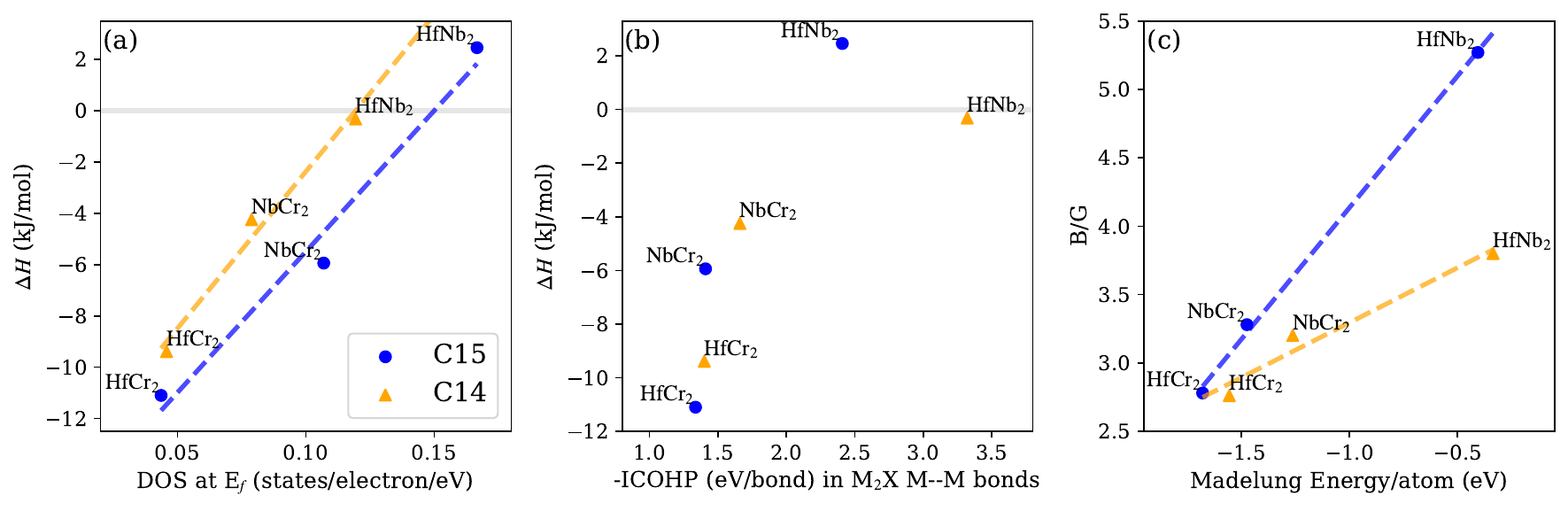}
    \caption{Plots of the enthalpy of formation as a function of (a) the total density of states at the Fermi level and (b) the negative value of the integral of the COHP up to the Fermi energy (-ICOHP) for M--M bonds. (c) Correlation between the Pugh's ratio ($B/G$) and the Madelung electrostatic potential energy per atom. Blue and yellow indicates C15 and C14 Laves phases, respectively.}
    \label{fig:comparisons}
\end{figure*}

The projected density of states (DOS) for C15 and C14 Laves phase stable compounds and for HfNb$_2$--C15 (positive enthalpy) is presented
in Fig. \ref{fig:DOS}. From the electronic DOS at the Fermi
level (E$F$), it is clear that all these XM$_2$ compounds exhibit
metallic behavior. The DOS near the Fermi levels for all compounds have two regions of peaks in DOS ($1.5$ up to $2$ states/electron/eV) separated by a low-density valley region ($0.4$ up to $0.6$ states/electron/eV). XCr$_2$ compounds have the region at E$_F$ dominated by Cr states (Fig. \hyperref[fig:DOS]{~\ref*{fig:DOS}(a)-(b), (d)-(e)}), and HfNb$_2$ compounds have a more balanced distribution (Fig. \hyperref[fig:DOS]{~\ref*{fig:DOS}(c), (f)}). Comparing the DOS at E$_F$ with the formation enthalpies, as shown in Fig. \hyperref[fig:comparisons]{~\ref*{fig:comparisons}(a)}, we can verify its influence in the destabilization of the C14 and C15 phases: HfNb$_2$--C15, with a DOS at E$_F$ of 0.167 states/electron/eV and a enthalpy of $2.46$ kJ/mol; HfNb$_2$--C14, with a DOS at E$_F$ of $0.12$ states/electron/eV and the least negative value of enthalpy, $-0.32$ kJ/mol.

In order to elucidate the stabilization mechanism of the XM$_2$ C15 and C14 compounds, we analysed four and five different bond types in C15 and C14 Laves phases, respectively. The crystal orbital Hamilton population (COHP) allow us to determine whether these bonds have bonding, anti-bonding or non-bonding interactions, in which its values are negative, positive and zero, respectively \cite{COHP_PWBS}. A measure of the covalent-bond strength (stronger bonds have a more negative value) can be made integrating the COHP with respect to energy within the valence bands (ICOHP), thus indicating the magnitude of net energy gain due to the bond interaction. The negative of the COHP (-COHP) curves of C15 M--M (atoms in the 16d site) bonds and C14 M--M (atoms in 2a and 6h sites) bonds are depicted in Fig. \ref{fig:COHP}, and our calculated ICOHP values are summarized in Table \ref{ICOHP}. Comparing the ICOHP values for M--M bonds in each compound, it is notable that they are relatively strong compared to the other bonds, thus having a notable influence into the structures stability \cite{cryst8050225}. From the M--M COHP curves, we can observe that bonding behavior dominates in the valence region, stabilizing the structures, but around -1 eV before E$_F$, anti-bonding character appears. The peak before the low-density valley region in
the electronic DOS at E$_F$ corresponds to the
separating boundary between the bonding and anti-bonding states. All six compounds have anti-bonding interacions at E$_F$ in M--M bonds, but HfNb$_2$ in both C15 and C14 Laves phases have more than the double of the -COHP value (indicated by the red dot in the plot) compared to the other compounds. The relationship between M--M bond type ICOHP values and the formation enthalpy for each compound, depicted in Fig. \hyperref[fig:comparisons]{~\ref*{fig:comparisons}(b)}, which does not follows a linear tendency, contributes to the idea that M--M bonds in HfNb$_2$-C15 act as a destabilization mechanism. 

A direct relation between ICOHP values for a compound and its mechanical properties can be stablished. For example, in C15 $(100)$ and $(111)$ planes, where the unique bonding interaction is X--X, we can directly observe that stronger interactions occurs in NbCr$_2$ and HfCr$_2$ than in HfNb$_2$ (ICOHP values of - 0.034, -0.023 and -0.007 eV/bond, respectively)  increasing the directional Young modulus.
In C15 plane $(110)$, the combination of M--X and X1--X1 bonds form a linear chain in a direction near $\frac{\pi}{3}$ rad, as seen in Fig. \hyperref[fig:comparisons]{\ref{fig:CHG}(b)}, representing the global Young modulus maximum.  For C14, planes $(001)$ are isotropic, as expected, following the same trend of C15 $(111)$ with similar values of B and E and ICOHP. The plane C14 $(100)$ has two different bonds: M1--M1 and M1--M2, the first alongside $[001]$ and the other in $[010]$ direction, respectively. M1--M2 bonds are shorter and have more strength than M1--M1, thus the C14 Young modulus highest value in this plane are in the $[010]$ direction. In the $(\bar{1}\bar{1}0)$ plane of C14, we can understand how the presence of X atoms influence the strength of the compound, as direction $[\bar{1} \bar{1} 0]$, with highest Young modulus, is crossed by X--X and M--X bonds.

Post-processing calculations of the charge density minus the superposition of atomic densities ($\Delta\rho_*$) were performed for the NbCr$_2$--C15 and HfNb$_2$--C14 compounds. This analysis, illustrated in three planes for each compound in Fig. \ref{fig:CHG}, is appropriate for understanding the charge redistribution in these phases. Positive and negative values of $\Delta\rho_*$ indicate regions of charge accumulation and depletion, respectively. In the plane $(110)$ of C15, represented in Fig. \hyperref[fig:comparisons]{\ref{fig:CHG}(b)}, there are tunnels of charge transfer between Nb and Cr (M--X, ICOHP of $-1.099$ eV) and between Nb and Nb (X1--X1, ICOHP of $-1.421$). Also, in the plane $(\bar{1}\bar{1}0)$ of C14 (Fig. \hyperref[fig:comparisons]{\ref{fig:CHG}(f)}), we can see that the M atoms in 2a and 6h wyckoff sites, which are surrounded by a negative value of $\Delta\rho_*$, have directional charge transfer to X atoms in 4f sites, but do not present direct charge transfer in any M--M bonds. In both compounds we can verify the metallic behavior with delocalized bonding, represented by the accumulation of charge in the interatomic region.

To have a deep understanding of the bonding behavior and ionicity in C15 and C14 Laves phases, we computed the Madelung electrostatic potential energy from the wave-function-based Mulliken charges, as implemented in LOBSTER \cite{MADELUNG_LOBSTER}, using the Bunge \cite{BUNGE1993113} auxiliary local basis set for projection. The correlation between Pugh's ratio and the Madelung electrostatic potential for C15 and C14 compounds, as shown in Fig.~\ref{fig:comparisons}(c), indicates that HfNb$_2$ are more ductile than XCr$_2$ (X = Nb, Hf) compounds because the former exhibits less charge transfer and thus a more metallic behavior, which is an expected trend based on the differences in electronegativity. This correlation of softening, Madelung energy and destabilization is similar to the observed in bcc Transition Metals \cite{PhysRevLett.103.235501}. The Mulliken charges were also calculated for C15 and C14 compounds (see
Tables S1 and S3 in the Supplemental Material).

\section{Conclusions}

\textit{Ab initio} calculations were performed to investigate the structural, elastic, anisotropic, and electronic properties, as well as the bonding character of the Laves phases in the Cr--Hf--Nb system in different configurations. The formation enthalpy showed that the HfNb$_2$-C14 compound has a negative value, indicating that it is thermodynamically stable at low temperatures. The elastic properties revealed that HfNb$_2$ is mechanically stable at zero pressure. The calculated elastic moduli for HfNb$_2$-C14 were lower than those of NbCr$_2$-C14 and HfCr$_2$-C14, showing more ductile and brittle behavior. Furthermore, the Debye temperature obtained for HfNb$_2$-C14 was lower than that of NbCr$_2$-C14 and HfCr$_2$--C14, showing reduced thermal capacity. Finally, the calculated elastic anisotropy evidences that HfNb$_2$-C14 shows an anisotropic elastic behavior. Electronic structure analyses, based on DOS and ICOHP, reveal that the M–M bonds in HfNb$_2$–C14 have a more metallic character than those in NbCr$_2$–C14 and HfCr$_2$–C14, due to the smaller charge transfer between Hf and Nb atoms. However, the strongly antibonding components observed at the Fermi level for these bonds suggest a destabilization mechanism similar to that in HfNb$_2$–C15, where the –COHP values are nearly twice as high and correlate with its positive formation enthalpy. This indicates that the partial occupation of M–M antibonding states contributes to the reduction of thermodynamic stability in both phases, although more pronouncedly in the C15 structure. Consequently, HfNb$_2$-C14 can be considered a metastable phase: it possesses sufficiently low energy to exist (–0.32 kJ/mol), yet electronic destabilization via strongly antibonding M–M interactions prevents it from achieving the full stability observed in NbCr$_2$–C14 and HfCr$_2$-C14.

\section*{Declaration of interests}

The authors declare that they have no known competing financial interests or personal relationships that could have appeared to influence the work reported in this paper.

\section*{Acknowledgement}

The research was carried out using high-performance computing resources made available by the Superintendência de Tecnologia da Informação (STI), University of São Paulo. This study was financed in part by the Cooperation Group of Brazilian Universities (GCUB) International Mobility Program (Mob) - Coordenação de Aperfeiçoamento de Pessoal de Nível Superior - Brasil (CAPES). The authors acknowledge the financial support from the National Council for Scientific and Technological Development (CNPq), Brazil, grant number 311756/2022-0. This study was financed, in part, by the São Paulo Research Foundation (FAPESP), Brasil. Process Number 2024/21634-2. 

\section*{Supplementary}

\url{https://www.overleaf.com/read/vspggfyhsrmf#24889f}

\balance

\bibliographystyle{elsarticle-num}
\bibliography{referencias}

\end{document}